\documentclass{kapproc} % Computer Modern font calls
\usepackage{epsfig}

\kluwerbib
\normallatexbib

\begin{document}

\articletitle{Review of QGP signatures - ideas versus observables
\thanks{Supported by DFG, BMBF, GSI}}

%\articlesubtitle{This is an Article Subtitle}

\author{E. L. Bratkovskaya,
M. Bleicher, A. Dumitru, K. Paech, M. Reiter, S.~Soff, H.~St\"ocker, H.~Weber}

 \affil{Institut f\"{u}r Theoretische Physik,
   Universit\"{a}t Frankfurt, 60054 Frankfurt, Germany}
%\email{firstauthor@myuniv.edu}

\author{M. van Leeuwen}
 \affil{NIKHEF, Amsterdam, Netherlands}

\author{W. Cassing}
 \affil{Institut f\"{u}r Theoretische Physik,
   Universit\"{a}t Giessen, 35392 Giessen, Germany}

\begin{abstract}
We investigate  hadron production and
transverse hadron spectra in nucleus-nucleus collisions from 2
$A\cdot$GeV to 21.3 $A\cdot$TeV within two independent transport
approaches (UrQMD and HSD) based on quark, diquark, string and
hadronic degrees of freedom.
The enhancement of pion production in central Au+Au (Pb+Pb)
collisions relative to scaled $pp$ collisions (the 'kink')
is described well by both approaches without involving a phase transition.
However, the maximum in the $K^+/\pi^+$ ratio at 20 to 30 A$\cdot$GeV
(the 'horn') is missed by $\sim$ 40\%.
Also, at energies above $\sim$ 5 A$\cdot$GeV, the measured $K^{\pm}$
$m_{T}$-spectra have a larger inverse slope than expected
from the models.
Thus the pressure generated by hadronic interactions in the transport
models at high energies is too low.
This finding suggests that the additional pressure -
as expected from lattice QCD at finite quark chemical
potential and temperature - might be generated by strong interactions in the
early pre-hadronic/partonic phase of central heavy-ion collisions.
Finally, we discuss the emergence of density perturbations in a
first-order phase transition and why they might affect relative hadron
multiplicities, collective flow, and hadron mean-free paths at decoupling.
A minimum in the collective flow $v_2$ excitation function was discovered
experimentally at 40 A$\cdot$GeV - such a behavior has been predicted
long ago as signature for a first order phase transition.
\end{abstract}

\section{Introduction}

The phase transition from partonic degrees of freedom (quarks and
gluons) to interacting hadrons is a central topic of modern high-energy
physics. In order to understand the dynamics and relevant scales of
this transition laboratory experiments under controlled conditions are
presently performed with ultra-relativistic nucleus-nucleus collisions.
Had\-ro\-nic spectra and relative hadron abundancies from these experiments
reflect  important aspects of the dynamics in the hot and dense zone
formed in the early phase of the reaction.  Furthermore, as has been
proposed early by Rafelski and M\"uller \cite{Rafelski} the strangeness
degree of freedom might play an important role in distinguishing
hadronic and partonic dynamics.

In fact, estimates based on the Bjorken formula \cite{bjorken} for the
energy density achieved in central Au+Au collisions suggest that the
critical energy density for the formation of a quark-gluon plasma (QGP)
is by far exceeded during a few fm/c in the initial phase of the
collision at Relativistic Heavy Ion Collider (RHIC) energies
\cite{QM01}, but sufficient energy densities ($\sim$ 0.7-1 GeV/fm$^3$
\cite{Karsch}) might already be achieved at Alternating Gradient
Synchrotron (AGS) energies of $\sim$ 10 $A\cdot$GeV \cite{HORST,exita}.
More recently, lattice QCD calculations at finite temperature and quark
chemical potential $\mu_q$ \cite{Fodor} show a rapid increase of the
thermodynamic pressure $P$ with temperature above the critical
temperature $T_c$ for a phase transition to the QGP. The crucial
question is, however, at what bombarding energies the conditions  for
the phase transition (or cross-over)  might be fulfilled.

Presently,  transverse mass (or momentum) spectra of hadrons are in the
center of interest. It is experimentally observed that the
transverse mass  spectra of kaons at AGS and SPS energies show a
substantial {\it flattening} or {\it hardening}  in central Au+Au
collisions relative to $pp$ interactions (cf.
Refs.~\cite{NA49_T,Goren}).  In order to quantify this effect, the
spectra are often parametrised as:
\begin{eqnarray}
\label{slope}
\frac{1}{m_T} \frac{dN}{dm_T} \sim \exp(-\frac{m_T}{T})
\end{eqnarray}
where $m_T=\sqrt{m^2+p_T^2}$ is the transverse mass and $T$ is the
inverse slope parameter. This hardening of the spectra is commonly
attributed to strong collective flow, which is absent in the
$pp$ or $pA$ data.

The authors of Refs. \cite{SMES} have proposed to interpret the
approximately constant $K^\pm$ slopes above $\sim 30$ A$\cdot$GeV -- the
'step' -- as an indication for a phase transition
following an early suggestion by Van Hove \cite{Hove}.  This
interpretation is also based on a rather sharp maximum in the
$K^+/\pi^+$ ratio at $\sim$ 20 to 30 A$\cdot$GeV in central
Au+Au (Pb+Pb) collisions (the 'horn' \cite{SMES}).  However,
it is presently not clear, if the statistical model assumptions invoked
in Refs. \cite{SMES} hold to be reliable.

We will demonstrate in this contribution that the pressure needed to
generate a large collective flow -- to explain the hard slopes of the
$K^\pm$ spectra as well as the 'horn' in the $K^+/\pi^+$ ratio -- is not
produced in the present models by the interactions of hadrons in the
expansion phase of the hadronic fireball.  In our studies we use two
independent transport models that employ hadronic and string degrees of
freedom, i.e., UrQMD (v. 1.3) \cite{UrQMD1,UrQMD2} and HSD
\cite{Geiss,Cass99}. They take into account the formation and multiple
rescattering of hadrons and thus dynamically describe the generation of
pressure in the hadronic expansion phase. This involves also
interactions of 'leading' pre-hadrons that contain a valence quark
(antiquark) from a 'hard' collision (cf. Refs.  \cite{Geiss,Weber02}).

The UrQMD transport approach \cite{UrQMD1,UrQMD2} includes all baryonic
resonances up to  masses of 2 GeV as well as mesonic resonances up to
1.9 GeV as tabulated by the Particle Data Group \cite{PDG}. For
hadronic continuum excitations a string model is used with hadron
formation times in the order of 1-2~fm/c depending on the momentum and
energy of the created hadron.
In the HSD approach nucleons, $\Delta$'s, N$^*$(1440), N$^*$(1535),
$\Lambda$, $\Sigma$ and $\Sigma^*$ hyperons, $\Xi$'s, $\Xi^*$'s and
$\Omega$'s  as well as their antiparticles are included on the baryonic
side whereas the $0^-$ and $1^-$ octet states are included in the
mesonic sector. High energy inelastic hadron-hadron collisions in HSD are
described by the FRITIOF string model \cite{LUND} whereas low energy
hadron-hadron collisions are modeled based on experimental cross
sections. Both transport approaches reproduce the nucleon-nucleon,
meson-nucleon and meson-meson cross section data in a wide kinematic
range.  We point out, that no explicit parton-parton scattering
processes (beyond the interactions of 'leading' quarks/diquarks) are
included in the studies below contrary to the multi-phase transport
model (AMPT) \cite{Ko_AMPT}, which is currently employed from upper SPS
to RHIC energies.

\section{Hadron excitation functions and ratios}

\subsection{$p p$ versus central $A A$ reactions -- the 'kink'}

\begin{figure}[t]
\begin{center}
\begin{minipage}[l]{5cm}
\epsfig{file=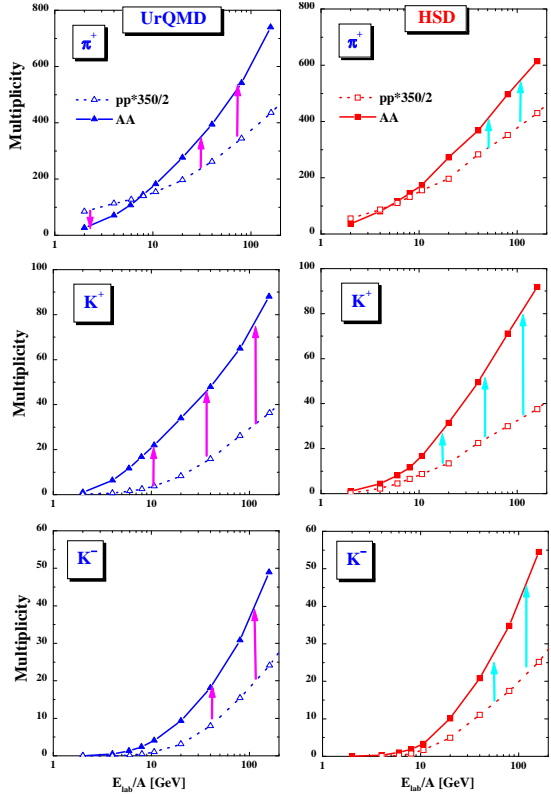,scale=0.4,clip}
\end{minipage}
\hfill\phantom{a}\begin{minipage}[l]{3.7cm}
\caption{Total multiplicities of $\pi^+, K^+$ and
$K^-$ (i.e., $4\pi$ yields) from central Au+Au (at AGS) or Pb+Pb
(at SPS) collisions in comparison to the total
multiplicities from $pp$ collisions (scaled by a factor 350/2)
versus kinetic energy $E_{\rm lab}$. The solid lines with full
triangles and squares show the UrQMD (l.h.s.) and HSD results
(r.h.s.) for $AA$ collisions, respectively. The dotted lines with
open triangles and squares correspond to the $pp$ multiplicities
calculated within UrQMD (l.h.s.) and HSD (r.h.s.).
The figure is taken from Ref. \protect\cite{Weber02}.}
\label{multppaa}
\end{minipage}
\end{center}
%\vspace*{-8mm}
\end{figure}

In order to explore the main physics from central $A A$ reactions
it is instructive to have a look at the various particle
multiplicities relative to scaled $pp$ collisions as a function of
bombarding energy. For this aim we show in Fig. \ref{multppaa}  the
total multiplicities of $\pi^+, K^+$ and $K^-$ (i.e., the $4\pi$
yields) from central Au+Au (at AGS) or Pb+Pb (at SPS)
collisions (from UrQMD and HSD) in comparison to the scaled
total multiplicities from $pp$
collisions versus the kinetic energy per particle $E_{\rm lab}$.

The general trend from both transport approaches is quite similar:
we observe a slight absorption of pions at lower bombarding energy
and a relative enhancement of pion production by rescattering
in heavy-ion collisions above $\sim$10 A$\cdot$GeV. Kaons and antikaons from
$AA$ collisions are always enhanced in central reactions relative
to scaled $pp$ multiplicities, which is a consequence of strong final
state interactions.  Thus, the 'kink' in the pion ratio as well as the
$K^\pm$ enhancement might result from conventional hadronic final state
interactions.

\subsection{Particle yields in central collisions of heavy nuclei}

\begin{figure}[t]
\begin{center}
\begin{minipage}[l]{8cm}
\epsfig{file=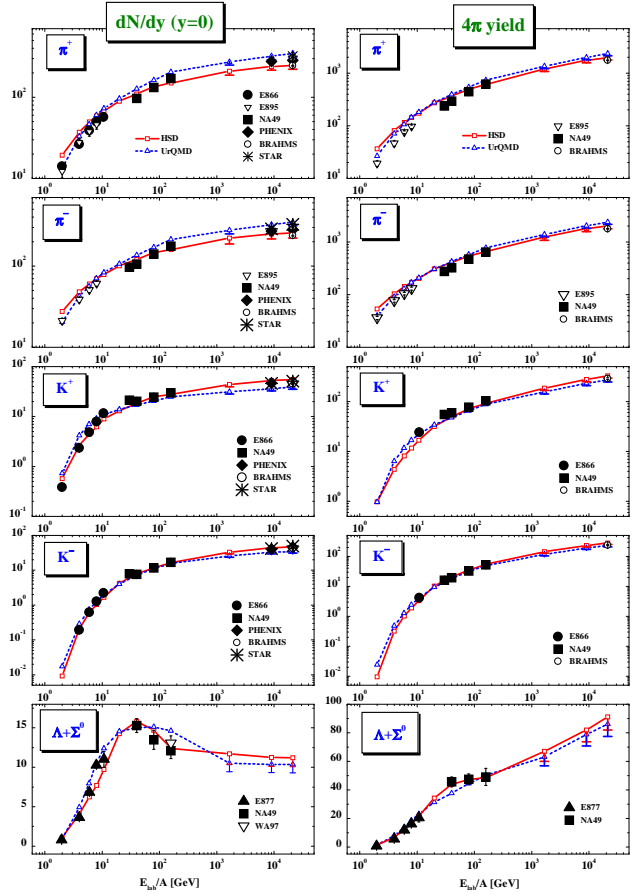,scale=0.44}
\end{minipage}
\hfill\phantom{a}\begin{minipage}[l]{3.1cm}
\caption{The excitation function of $\pi^+, \pi^-, K^+, K^-$ and
$\Lambda+\Sigma^0$ yields from central
Au+Au or Pb+Pb collisions in comparison to the
experimental data from Refs.  \protect\cite{E866E917,E895,E891Lam}
(AGS), \protect\cite{NA49_new,NA49_Lam,Antiori} (SPS)
and \protect\cite{BRAHMS,PHENIX,STAR} (RHIC) for midrapidity
(left column) and rapidity integrated yields (right column).  The solid
lines with open squares show the results from HSD whereas the dashed
lines with open triangles indicate the UrQMD calculations. The lower
theoretical errorbars at RHIC energies correspond to the yields for
10\% central events.
The figure is taken from Ref. \protect\cite{Bratnew}.}
\label{Fig_yield}
\end{minipage}
\vspace*{-18mm}
\end{center}
\end{figure}
Fig. \ref{Fig_yield} shows the excitation function of $\pi^+, \pi^-,
K^+, K^-$ and $\Lambda+\Sigma^0$ yields (midrapidity (l.h.s.) and rapidity
integrated (r.h.s)) from central Au+Au (Pb+Pb) collisions in
comparison to the experimental data.
Note that all data from the NA49 Collaboration at 30 A$\cdot$GeV
have to be considered as 'preliminary'.
As can be seen from Fig. \ref{Fig_yield} the differences between the
independent transport models are less than 20\%.  The maximum
deviations between the models and the experimental data are less than
$\sim 30$\%. In addition, a systematic analysis of the results from both
models and experimental data for central nucleus-nucleus collisions
from 2 to 160 $A\cdot$GeV in Ref. \cite{Weber02} has shown that also
the 'longitudinal' rapidity distributions of protons, pions, kaons,
antikaons and hyperons are quite similar in both models and in
reasonable agreement with available data.  The exception are the pion
rapidity spectra at the highest AGS energy and lower SPS energies,
which are overestimated by both models \cite{Weber02}.  For a more
detailed comparison of HSD and UrQMD calculations with experimental
data at RHIC energies we refer the reader to Refs.
\cite{Brat03,Soff03}.

\subsection{Particle ratios -- the 'horn'}

\begin{figure}[t]
\begin{center}
\begin{minipage}[l]{8cm}
\epsfig{file=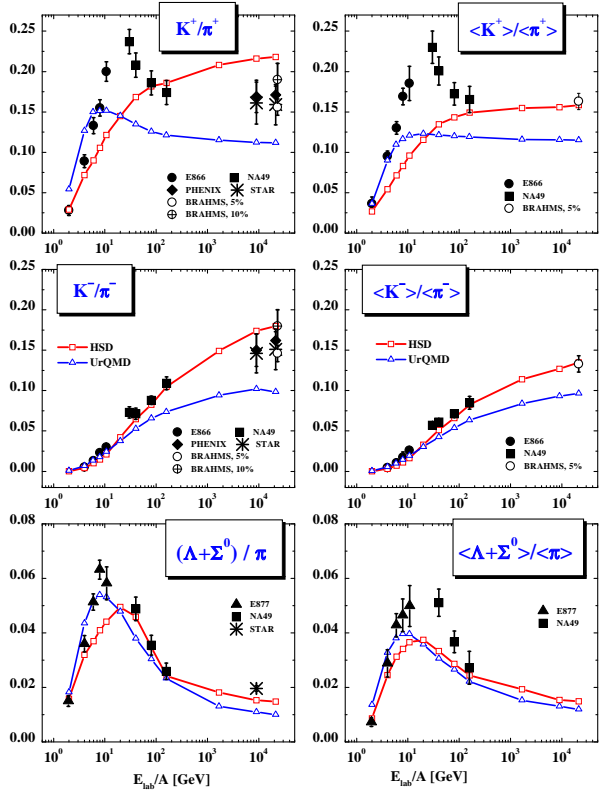,scale=0.42}
\end{minipage}
\phantom{a}\begin{minipage}[l]{3.5cm}
\caption{The excitation function of $K^+/\pi^+, K^-/\pi^-$ and
$(\Lambda+\Sigma^0)/\pi$ ratios from 5\% central (AGS energies, SPS at 160
A$\cdot$GeV and at RHIC energies), 7\% central (20, 30, 40 and 80
A$\cdot$GeV), 10\% central for $\Lambda+\Sigma^0$ at 160 A$\cdot$GeV
Au+Au (AGS and RHIC) or Pb+Pb (SPS) collisions in comparison to the
experimental data from Refs. \protect\cite{E866E917,E891Lam}
(AGS), \protect\cite{NA49_new,NA49_Lam,Antiori} (SPS) and
\protect\cite{BRAHMS,PHENIX,STAR} (RHIC) for midrapidity (left column)
and rapidity integrated yields (right column).  The solid lines with
open squares show the results from HSD whereas the dashed lines with
open triangles indicate the UrQMD calculations.
The figure is taken from Ref. \protect\cite{Bratnew}.}
\label{Fig_rat}
\end{minipage}
\end{center}
\vspace*{-8mm}
\end{figure}
In Fig. \ref{Fig_rat} we present the excitation function of the
particle ratios $K^+/\pi^+, K^-/\pi^-$ and $(\Lambda+\Sigma^0)/\pi$
from central Au+Au (Pb+Pb) collisions in comparison to
experimental data.
The deviations between the transport models and the data are most
pronounced for the midrapidity ratios (left column) since the ratios
are very sensitive to actual rapidity spectra. The $K^+/\pi^+$ ratio in
UrQMD shows a maximum at $\sim$ 8 A$\cdot$GeV and then drops to a
constant ratio of 0.11 at top SPS and RHIC energies. In the case of HSD
a continuously rising ratio with bombarding energy is found for the
midrapidity ratios which partly is due to a dip in the pion
pseudo-rapidity distribution at RHIC energies (cf. Fig. 1 in Ref.
\cite{Brat03}). The 4$\pi$ ratio in HSD is roughly constant  from top
SPS to RHIC energies, however, larger than the ratio from UrQMD due to
the lower amount of pion production in HSD essentially due to an
energy-density cut of 1 GeV/fm$^3$, which does not allow to form
hadrons above this critical energy density \cite{Weber02} and a
slightly higher $K^+$ yield (cf. Fig. \ref{Fig_yield}).  Nevertheless,
the experimental maximum in the $K^+/\pi^+$ ratio is missed,  which we
address dominantly to the excess of pions in the transport codes rather
than to missing strangeness production.  Qualitatively, the same
arguments - due to strangeness conservation - also hold for the
$(\Lambda +\Sigma^0)/\pi$ ratio, where the pronounced experimental
maxima are underestimated due to the excess of pions in the transport
models at top AGS energies (for HSD) and above $\sim$ 5 A$\cdot$GeV
(for UrQMD). Since the $K^-$ yields are well reproduced by both
approaches (cf. Fig. \ref{Fig_yield}) the deviations in the $K^-/\pi^-$
ratios at SPS and RHIC energies in UrQMD can be traced back to the
excess of pions (see discussion above).

We stress that the maximum in the $(\Lambda +\Sigma^0)/\pi$ ratio is
essentially due to a change from baryon to meson dominated dynamics
with increasing bombarding energy. Similar arguments hold for the
experimentally observed maxima in the ratio $\Xi/\pi$  (cf. Ref.
\cite{Xipi}). However, the 'horn' in the $K^+/\pi^+$ ratio at
$\sim$30 A$\cdot$GeV is not described by neither of our transport models.

\section{Transverse mass spectra -- the 'step'}

We now focus on transverse mass spectra of pions and kaons/antikaons
from central Au+Au (Pb+Pb) collisions from 2 $A\cdot$GeV to 21.3
$A\cdot$TeV  and compare to recent data (cf. Ref. \cite{Brat03PRL}).
Without explicit representation we mention that the agreement between
the transport calculations and the data for $pp$ and for central C+C and
Si+Si is quite satisfactory \cite{Brat03PRL}; no obvious traces of
'new' physics are visible.  The situation, however, changes for central
Au+Au (or Pb+Pb) collisions.  Whereas at the lowest energy of 4
$A\cdot$GeV the agreement between the transport approaches and the data
is still acceptable, severe deviations are visible in the $K^\pm$
spectra at SPS energies of 30 and 160 $A\cdot$GeV \cite{Brat03PRL}.  We
note that the $\pi^{\pm}$ spectra are reasonably described at all
energies while the inverse slope $T$ of the $K^\pm$ transverse mass
spectra in Eq. (\ref{slope}) is underestimated severely by about the
same amount in both transport approaches (within statistics).  The
increase of the inverse $K^\pm$ slopes in heavy-ion collisions with
respect to $pp$ collisions, which is generated by rescatterings of
produced hadrons in the transport models, is small because the
elastic meson-baryon scattering is strongly forward peaked and
therefore gives little additional transverse momentum at midrapidity.
\begin{figure}[t]
\begin{center}
\begin{minipage}[l]{8.1cm}
\epsfig{file=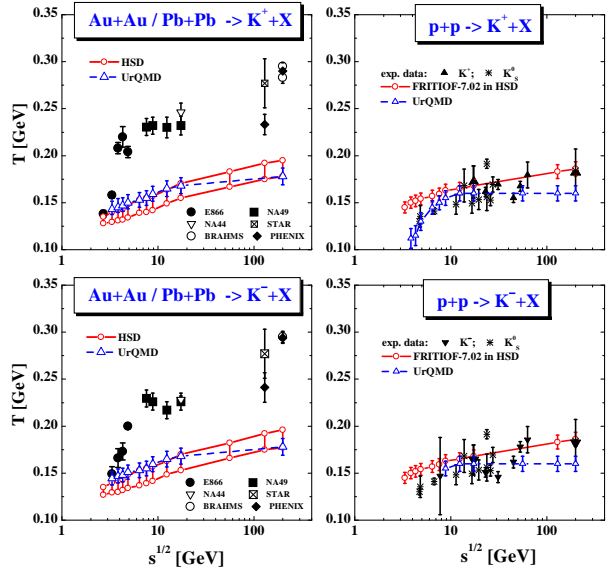,scale=0.41}
\end{minipage}
\begin{minipage}[l]{3.4cm}
\caption{Comparison of the inverse slope parameters $T$ for $K^+$ and
$K^-$ mesons from central Au+Au (Pb+Pb) collisions (l.h.s.) and $pp$
reactions (r.h.s.) as a function of the invariant energy $\sqrt{s}$ from
HSD (upper and lower solid lines) and UrQMD (open triangles) with
data from Refs. \protect\cite{E866E917,NA49_T,NA44,STAR,BRAHMS,PHENIX}
for $AA$ and \protect\cite{NA49_CCSi,Gazdz_pp,STAR} for $pp$ collisions.
The upper and lower solid lines result from different limits of
the HSD calculations as discussed in the text.
The figure is taken from Ref. \protect\cite{Bratnew}.}
 \label{Fig_T}
\end{minipage}
\end{center}
\vspace*{-8mm}
\end{figure}

The question remains whether the underestimation of the $K^\pm$ slopes
in the transverse mass spectra \cite{Brat03PRL} might be due to
conventional hadronic medium effects.  In fact, the $m_T$ slopes of
kaons and antikaons at SIS energies (1.5 to 2 $A\cdot$GeV) were found
to differ significantly \cite{KaoS}. As argued in  \cite{Cass99}
the different slopes could be traced back to repulsive kaon-nucleon
potentials, which lead to a hardening of the $K^+$ spectra,
and attractive antikaon-nucleon potentials, which lead to a softening
of the $K^-$ spectra. However, the effect of such potentials was
calculated within HSD and  found to be of minor importance at AGS and
SPS energies \cite{Cass99} since the meson densities are comparable to
or even larger than the baryon densities at AGS energies and above.
Additional self energy contributions stem from $K^\pm$ interactions
with mesons; however, $s$-wave kaon-pion interactions are weak due to
chiral symmetry arguments and $p$-wave interactions such as $\pi+K
\leftrightarrow K^*$ transitions are  suppressed substantially by the
approximately 'thermal' pion spectrum \cite{Fuchs}.

Furthermore, we have pursued the idea of Refs.  \cite{Sorge,Bleich}
that the $K^\pm$ spectra could be hardened by string-string
interactions, which increase the effective string tension $\sigma$ and
thus the probability to produce mesons at high $m_T$
\cite{Ko_AMPT,Bleich}.  In order to estimate the largest possible
effect of string-string interactions we have assumed that for two
overlapping strings the string tension $\sigma$ is increased by a
factor of two, for three overlapping strings by a factor of three etc.
Here the overlap of strings is defined geometrically assuming a
transverse string radius $R_s$, which according to the studies in Ref.
\cite{Geiss99} should be $R_s \leq$ 0.25 fm.  Based on these
assumptions (and $R_s$=0.25 fm), we find only a small increase of the
inverse slope parameters at AGS energies, where the string densities
are low.  At 160 $A\cdot$GeV the model gives a  hardening of
the spectra by about 15\%, which, however, is still significantly less
than the effect observed in the data.

Our findings are summarized in Fig. \ref{Fig_T}, where the dependence of
the inverse slope parameter $T$ (see Eq.~(\ref{slope})) on $\sqrt{s}$
is shown and compared to the experimental data \cite{NA49_T,NA49_CCSi} for
central Au+Au (Pb+Pb) collisions (l.h.s.) and $pp$ reactions (r.h.s.).
The upper and lower solid lines (with open circles) on the l.h.s. in
Fig. \ref{Fig_T} correspond to results from HSD calculations, where the
upper and lower limits are due to fitting the slope $T$ itself, an
uncertainty in the repulsive $K^\pm$-pion potential or the possible
effect of string overlaps.
The slope parameters from $pp$ collisions (r.h.s. in Fig.  \ref{Fig_T})
are seen to increase smoothly with energy both in the experiment (full
squares) and in the HSD calculations (full lines with open circles).
The UrQMD results for $pp$ collisions are shown as open triangles
connected by the solid line and systematically lower than the slopes
from HSD at all energies.

We mention that the RQMD model \cite{Sorge} gives higher inverse slope
parameters for kaons at AGS and SPS energies than HSD and UrQMD, which
essentially might be traced back to the implementation of effective
resonances with masses above 2 GeV as well as 'color ropes' which decay
isotropically in their rest frame \cite{Hecke}.  A more detailed
discussion of this issue is presented in Ref.\cite{Bratnew}.

\section{Proton elliptic flow - the minimum}

\begin{figure}[h]
\phantom{a}\vspace*{-3mm}
\begin{center}
\begin{minipage}[l]{8cm}
\epsfig{file=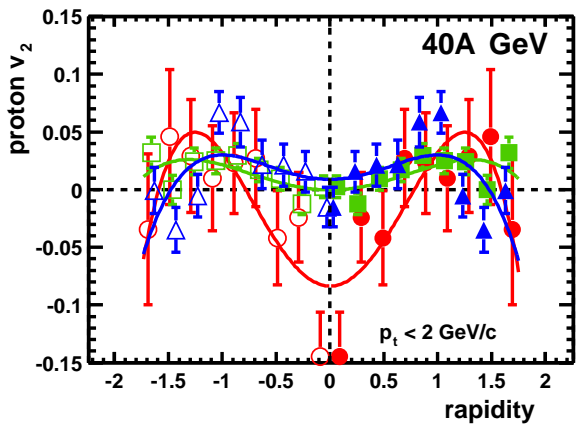,scale=1.3}
\end{minipage}
\begin{minipage}[l]{3.1cm}
\caption{Elliptic flow $v_2$ of protons versus rapidity from 40
A$\cdot$GeV Pb+Pb collisions \protect\cite{NA49_v2pr40} measured for
three centrality bins:  central (dots), mid-central (squares) and
peripheral (triangles).  The solid lines are polynomial fits to the
data \protect\cite{NA49_v2pr40}.}
\label{Fig_v2pr40}
\end{minipage}
\end{center}
\vspace*{-5mm}
\end{figure}

The NA49 Collaboration \cite{NA49_v2pr40} has recently observed
a vanishing elliptic flow of protons in Pb+Pb collisions at 40
A$\cdot$GeV at midrapidity for all centralities (Fig.
\ref{Fig_v2pr40}).  This observation of the apparent collapse of the
collective flow $v_2$ is remarkable because the proton
elliptic flow $v_2$ at top AGS (11 A$\cdot$GeV) \cite{E895_00v2} and
top SPS energies (160 A$\cdot$GeV) \cite{NA49_v2pr40} is
non-zero for mid-central collisions and large for peripheral
collisions. This experimental observation of a minimum of the collective
flow excitation function has been predicted as a signature for a first
order phase transition \cite{Horst76}.

\section{Thermodynamics in the $T-\mu_B$ plane}

This still leaves us with the question of the origin of the rapid
increase of the $K^\pm$ slopes with invariant energy for central Au+Au
collisions at AGS energies and the constant slope at SPS energies (the
'step'), which is  missed in both transport approaches. We recall that
higher transverse particle momenta either arise from repulsive self
energies -- in mean-field dynamics -- or from collisions, which reduce
longitudinal momenta in favor of transverse momenta \cite{HORST,CaMo}.
As shown above in Fig. \ref{Fig_T} conventional hadron self-energy
effects and hadronic binary collisions are insufficient to describe the
dramatic increase of the $K^\pm$ slopes as a function of $\sqrt{s}$.
This indicates additional mechanisms for the generation of the pressure
that is observed experimentally.

\begin{figure}[t]
\centerline{\psfig{figure=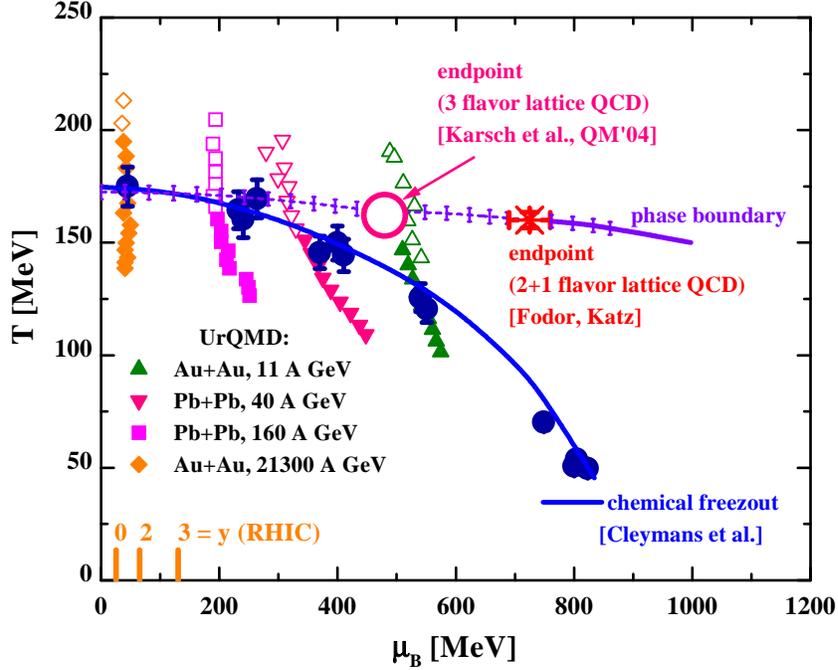,width=11cm}}
\caption{ Schematic phase diagram in the $T-\mu_B$ plane.
The solid line characterizes the universal chemical
freeze-out line from Cleymans et al. \protect\cite{Cleymans} whereas
the full dots (with errorbars) denote the 'experimental' chemical
freeze-out parameters from Ref. \protect\cite{Cleymans}. The various
symbols stand for temperatures $T$ and chemical potentials $\mu_B$
extracted from UrQMD transport calculations in central Au+Au (Pb+Pb)
collisions at 21.3 A$\cdot$TeV, 160, 40 and 11 A$\cdot$GeV
\protect\cite{Bravina} (see text). The stars indicate the tri-critical
endpoints from lattice QCD calculations by Karsch et al.
\protect\cite{Karsch2} (large open circle) and Fodor and Katz
\protect\cite{Fodor} (star). The 'horizontal' line with errorbars
is the phase boundary (from the endpoint) given in Ref. \protect\cite{Fodor}.
The 'vertical' lines indicate $\mu_B$ for different rapidity intervals
at RHIC energies from Ref. \protect\cite{BRAHMS_PRL03}.}
\label{Fig_QCD}
\end{figure}

Here we propose that additional pre-hadronic/partonic degrees of
freedom might be responsible for this effect already at $\sim$ 5
$A\cdot$GeV. Our arguments are based on a comparison of the
thermodynamic parameters $T$ and $\mu_B$ extracted from the transport
models in the central overlap regime of Au+Au collisions \cite{Bravina}
with the experimental systematics on chemical freeze-out configurations
\cite{Cleymans} in the $T,\mu_B$ plane. The solid line in Fig.
\ref{Fig_QCD} characterizes the universal chemical freeze-out line from
Cleymans et al. \cite{Cleymans} whereas the full dots with errorbars
denote the 'experimental' chemical freeze-out parameters - determined
from the fits to the experimental yields - taken from Ref.
\cite{Cleymans}. The various symbols (in vertical sequence) stand for
temperatures $T$ and chemical potentials $\mu_B$ extracted from UrQMD
transport calculations in central Au+Au (Pb+Pb) collisions at 21.3
A$\cdot$TeV, 160, 40 and 11 A$\cdot$GeV \cite{Bravina} as a function of
the reaction time (from top to bottom).  The open symbols denote
nonequilibrium configurations and correspond to $T$ parameters
extracted from the transverse momentum distributions, whereas the full
symbols denote configurations in approximate pressure equilibrium in
longitudinal and transverse direction.

During the nonequilibrium phase (open symbols) the transport
calculations show much higher temperatures (or energy densities) than
the 'experimental' chemical freeze-out configurations at all bombarding
energies ($\geq$ 11 A$\cdot$GeV).  These numbers are also higher than
the tri-critical endpoints extracted from lattice QCD calculations by
Karsch et al.  \cite{Karsch2} (large open circle) and Fodor and Katz
\cite{Fodor} (star). Though the QCD lattice
calculations differ substantially in the value of $\mu_B$ for the
critical endpoint, the critical temperature $T_c$ is in the range of
160 MeV in both calculations, while the energy density is in the order
of 1 GeV/fm$^3$ or even below. Nevertheless, this diagram shows that at
RHIC energies one encounters more likely a cross-over between the
different phases when stepping down in temperature during the expansion
phase of the 'hot fireball'. This situation changes at lower SPS or AGS
(as well as new GSI SIS-300) energies, where for sufficiently large
chemical potentials $\mu_B$ the cross over should change to a first
order transition \cite{Shuryak}, i.e., beyond the tri-critical point in
the ($T,\mu_B$) plane.  Nevertheless, Fig. \ref{Fig_QCD} demonstrates
that the transport calculations show temperatures (energy densities)
well above the phase boundary (horizontal line with errorbars) in the
very early phase of the collisions, where hadronic interactions
practically yield no pressure, but pre-hadronic degrees of freedom
should do. This argument is in line with the studies on elliptic flow
at RHIC energies, that is underestimated by ~30\% at midrapidity in the
HSD approach for all centralities \cite{Brat03}. Only strong early
stage pre-hadronic interactions might cure this problem.

In Fig. \ref{Fig_QCD} we also show the baryon chemical potential
$\mu_B$ for different rapidity intervals at RHIC energies as obtained
from a statistical model analysis by the BRAHMS Collaboration based on
measured the antihadron to hadron yield ratios \cite{BRAHMS_PRL03}.
For midrapidity, $\mu_B\simeq 0$, whereas for forward rapidities
$\mu_B$ increases up to $\mu_B\simeq 130$~MeV at $y=3$.  Thus, the
forward rapidity measurement allows to probe large $\mu_B$ at the same
bombarding energies.  Hence, at RHIC only a rather limited chemical
potential range is accessible experimentally. To reach the probable
first order phase transition region, the International Facility at GSI
seems to be the right place to go.

\section{Density perturbations from dynamical symmetry breaking}

It is of great interest, of course, to investigate whether the
above-mentioned observations could be due to a phase transition of
strongly interacting matter. The natural effective theory for exploring
the effects from phase transitions on the production and phase-space
distribution of hadrons is hydrodynamics: the equation of state enters
directly by closing the system of continuity equations for energy,
momentum and charge conservation. Typically, first-order phase
transitions are modelled by matching the pressure of the low-density
massive (symmetry broken) phase to that of the high-density massless
(symmetric) phase along a `phase boundary' in the $T-\mu_B$ plane
(cf. Fig. \ref{Fig_QCD}). On the phase transition
line, the system is in a mixed state where both phases coexist and
where their relative fractions are determined from Gibbs's conditions
of phase equilibrium. This construction assumes that the phase
transition is a quasi-static, reversible process (entropy is conserved)
near equilibrium. Entropy is produced only in the initial compression
stage which ends with the formation of a locally equilibrated fireball
of hot and dense matter which subsequently expands and cools.

The crossover from suppressed to increased pion production in central
nuclear collisions relative to scaled $pp$ collisions reflects the
excess entropy produced at higher energies~\cite{Reiter:1998uq}, as
also seen in Fig.~1.  Somewhat surprisingly though, the excitation
function of entropy production turns out to be rather smooth, without
exhibiting `discontinuities' from crossing the phase boundary at some
energy. Aside from some dynamical effects, the main reason for this
behavior is that in {\em baryon-rich} matter the specific entropy is a
smooth function of temperature without a pronounced
`jump'~\cite{Reiter:1998uq}.

Since the entropy produced right at impact (on a time scale of order
$2R_A/\gamma$ in the CM frame) increases smoothly with energy, all
hadron abundance ratios will behave correspondingly and the sharp
`horn' in $K^+ / \pi^+$ seen in the data can not be reproduced. This
holds for typical hydro models with a first-order phase
transition~\cite{Reiter:1998uq} as well as for hadronic transport
models (see discussion above).

If early-stage entropy poduction can not account for the sharp peak of
$K^+ /\pi^+$ ratio then perhaps the phase transition {\em back} to the
broken phase (which occurs later on after some cooling) can~? This might
be possible indeed if one abandons the equilibrium phase transition based
on the macroscopic Gibbs construction and, in turn, introduces a
dynamical microscopic treatment of phase transitions into
hydrodynamics. It is well-known that first-order phase transitions lead
to inhomogeneities such as high-density `nuggets', surrounded by
low-density `voids'. Analogous effects are frequently discussed within
the context of the QCD transition in the early universe, where
inhomogeneities of the entropy (or baryon to photon ratio) might affect
BBN.  The usual mixed-phase construction applies on scales much larger
than the size and separation of inhomogeneities, and on such scales the
matter and entropy distributions appear smooth. On small scales
however, for example in heavy-ion collisions, inhomogeneous density
distributions have significant effects on observables which are
non-linear functions of the density: take a large homogeneous system,
split it in half, and move all baryons into one half, then let each
half equilibrate. Suppose you can not measure the hadron multiplicities
in each half separately, just the total.  The obvious measurement,
namely of baryon number, doesn't reflect the presence of the
high-density nugget because the total baryon number is the same as for
the homogeneous distribution. However, the total yield of $K^+$ over
the total yield of $\pi^+$ will be larger than for the homogeneous
system~! This is because the ratio is enhanced in the high-density
nugget by a much bigger factor than it is suppressed in the low-density
half of the system. (Other hadrons like (multi-)strange baryons of
course compensate the strangeness and are also more abundant than for
the homogeneous system.) The effect diminishes rapidly when the entropy
per (net) baryon becomes large, that is, in the meson-dominated
high-energy regime.
\begin{figure}[t]
\begin{center}
\begin{minipage}[l]{6cm}
\epsfig{file=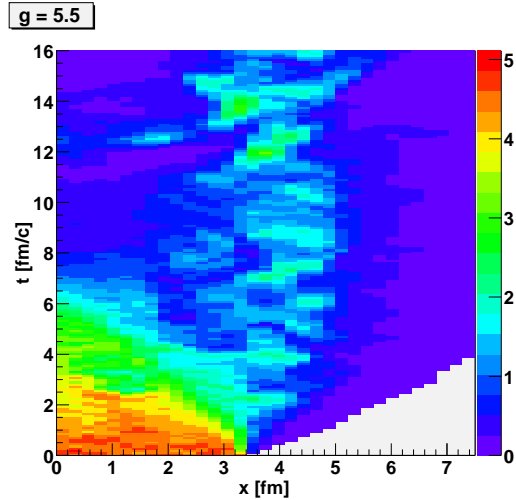,scale=0.35,clip}
\end{minipage}
\hfill\phantom{a}\begin{minipage}[l]{4cm}
\caption{Space-time evolution of the fluid energy density along the
$x$-axis at $y=z=0$. The scale on the right specifies the ener\-gy density
in units of nuc\-lear matter density $\epsilon_0\simeq 150$ MeV/fm$^3$.}
\label{edens_chhyd}
\end{minipage}
\end{center}
\vspace*{-8mm}
\end{figure}
To investigate the formation of inhomogeneities during the phase
transition we solve for the coupled evolution of an order parameter
field such as the chiral condensate $\phi$ and the thermalized matter
fields~\cite{Paech:2003fe}:
\begin{eqnarray}
\fbox{\phantom{a}} \ \phi + \partial V_{\rm eff}/\partial\phi =0, \ \ \
\partial_\mu \left( T_{\rm fl}^{\mu\nu} + T_\phi^{\mu\nu}\right)=0.
\label{hyd}\end{eqnarray}
Here, $T_{\rm fl}^{\mu\nu}$ is the energy-momentum tensor of the fluid,
$T_\phi^{\mu\nu}$ that of the classical modes of the chiral condensate,
and $V_{\rm eff}$ is the effective potential obtained by integrating
out the thermalized degrees of freedom.  We focus first on
energy-density inhomogeneities and present solutions of these coupled
equations for vanishing baryon density~\cite{Paech:2003fe}. As initial
condition we chose a homogeneous energy density above the critical
energy density for the transition to the broken phase. However, the
condensate $\phi$ exhibits `primordial' Gaussian fluctuations on length
scales $\sim 1$~fm on top of a smoothly varying mean field. These
fluctuations are then propagated through a first-order chiral phase
transition and leave a rather inhomogeneous (energy-) density
distribution in the wake of the transition, as seen in
Fig.~\ref{edens_chhyd}. Evidently, the scale for such fluctuations is
not tiny and so it would not be appropriate to assume a homogeneous
density distribution. On the other hand, they are too small to be
resolved in rapidity space because the scale factor is large at times
long after the initial impact. To resolve individual hot/dense spots
would require a resolution better than one unit of rapidity, which is
roughly equal to the thermal width of the local particle momentum
distributions.

However, additional hints for the existence of large density
inhomogeneities created in the course of the transition to the broken
phase remain to be explored.  (Inhomogeneities from fluctuations of
particle production in the primary nucleon-nucleon collisions should be
largely washed out until decoupling by hydrodynamic transport of matter
due to pressure gradients, see e.g.~\cite{Bleicher:wd}.) Clearly, the
yields of other hadron species depend non-linearly on the density as
well, and their behavior has to be tested for consistency.  Moreover,
coordinate-space fluctuations of the energy-momentum tensor of matter
produced by a phase transition are uncorrelated to the reaction plane
and therefore should act to reduce out-of-plane collective flow
($v_2/\langle p_t\rangle$) as compared to equilibrium hydrodynamics,
cf.\ the discussion in~\cite{Paech:2003fe}.  Finally,
Hanbury-Brown--Twiss correlations could provide valuable
coordinate-space information on the regions from which particles are
emitted.  In this regard, note the stunning result of
CERES~\cite{Adamova:2002ff} according to which pions decouple when
their mean-free path is $\sim 1$~fm. This is inconceivable in standard
equilibrium hydrodynamics without density perturbations because there
particles decouple only when their mean-free path exceeds the scale of
spatial homogeneity, which is about an order of magnitude
larger~\cite{Soff:2000eh}.  The CERES analysis indicates that density
(and perhaps velocity) gradients in coordinate space are 1/1~fm rather
than 1/10~fm.

\section{Conclusions}

Summarizing this contribution, we point out that baryon stopping
\cite{Weber_stop02} and hadron production in central Au+Au (or Pb+Pb)
collisions is quite well described in the independent transport
approaches HSD and UrQMD. Also the 'longitudinal' rapidity
distributions of protons, pions, kaons, antikaons and hyperons are
similar in both models and in  reasonable agreement with available
data. The exception are the pion rapidity spectra at the highest AGS
energy and lower SPS energies, which are overestimated by both models
\cite{Weber02}.  As a consequence the HSD and UrQMD transport
approaches underestimate the experimental maximum of the $K^+/\pi^+$
ratio ('horn') at $\sim$ 20 to 30 A$\cdot$GeV.
However, we point out that the maxima in the $K^+/\pi^+$ and
($\Lambda+\Sigma^0)/\pi$ ratios partly reflect a change from baryon
to meson dominated dynamics with increasing bombarding energy.

We have found that the inverse slope parameters $T$ for $K^\pm$ mesons
from the HSD and UrQMD transport models are practically independent of
system size from $pp$ up to central Pb+Pb collisions and show only a
slight increase with collision energy, but no 'step' in the $K^\pm$
transverse momentum slopes. The rapid increase of the inverse slope
parameters of kaons for collisions of heavy nuclei (Au+Au) found
experimentally in the AGS energy range, however, is not reproduced by
neither model (see Fig.~\ref{Fig_T}).  Since the pion transverse mass
spectra -- which are hardly effected by collective flow  -- are
described sufficiently well at all bombarding energies
\cite{Bratnew}, the failure has to be attributed to a lack of pressure.
This additional pressure should be
generated in the early phase of the collision, where the 'transverse'
energy densities in the transport approaches are higher than the
critical energy densities for a phase transition (or cross-over) to
the QGP. The
interesting finding of our analysis is, that pre-hadronic degrees of
freedom might already play a substantial role in central Au+Au collisions
at AGS energies above $\sim$~5~$A\cdot$GeV.
The more astonishing is the experimentally observed collapse of
the collective flow $v_2$ at 40 A$\cdot$GeV - such a behavior has
been predicted as signature for a first order phase transition long ago.

\end{document}